\documentclass[final]{elsart}
\usepackage[nomarkers]{endfloat}
\usepackage{graphicx,amssymb,amsmath}
\usepackage[square,comma,numbers,sort&compress]{natbib}
\usepackage{amsfonts}
\usepackage{epsf}
\usepackage{psfrag}
\usepackage[pagewise,running]{lineno}
\usepackage{graphicx}
\usepackage{graphics}
\usepackage{setspace}
\usepackage{longtable}
\usepackage{lscape}
\usepackage{rotating}
\usepackage{dcolumn}
\usepackage[thinspace,squaren]{mySIunits}

\def\ie{{\em i.e.}}
\def\pp{{polypropylene}}
\def\etc{{\em etc.}}
\def\fe{{\textsc{fe}}}
\def\dd{{\mathrm{d}}}
\def\dotted{\protect\mbox{${\mathinner{\cdotp\cdotp\cdotp\cdotp\cdotp\cdotp}}$}}  
\def\full{\protect\mbox{------}}
\def\kesik{\protect\mbox{-\, -\, -\, -}}

\journal{arXiv.org}%%%%%Scripta Materialia}
\begin{document}
\begin{frontmatter}
\title{Elastic properties of cellular polypropylene  films: Finite element simulations and their comparison with experiments}

\author{Enis Tuncer\corauthref{cor}},
\corauth[cor]{Corresponging author.}
\ead{enis.tuncer@physics.org}
\author{Michael Wegener}
\address{Applied Condensed-Matter Physics, Department of Physics, University of Potsdam, D-14469 Potsdam, Germany}

%\author{\firstname Enis \surname Tuncer}
%\email{enis.tuncer@physics.org}
%\author{\firstname Michael \surname Wegener} 

%\affiliation{Applied Condensed-Matter Physics, Department of Physics, University of Potsdam, D-14469 Potsdam Germany}

%\date{\today}
%\volume{}

\begin{abstract}
The Young's modulus of a two-dimensional truss-like structure is simulated by using the finite element method. A power-law expression is proposed for the effective Young's modulus of the system. The obtained numerical results are compared with the experimental data of the {\em anisotropic thin cellular polypropylene films}. %The experimental data have lower Young's modulus values then the foam data in the literature. This is explained by the internal structure of the films. %Valuable information regarding the micro-structure of the films are obtained from the comparisons. 
At high solid volume fractions ($>0.4$), the average shape of the cells are lateral, and their dimensions have around one-to-five ratio. As the samples are inflated further, volume fraction of the solid is decreased, the average shape approach a diamond-like structure with one-to-two ratio. In addition the effective Young's modulus of the system increases. It is concluded that valuable structural information can be obtained by analyzing the experimental data and the numerical simulations, which take into account the material's micro-structural information, simultaneously. 
\end{abstract}
%keywords{Foams, Cellular materials, homogenization, effective medium,  Young's modulus, micro-structural information}
%\pacs{02.70.Hm, 07.05.Kf, 61.18-j, 62.20.-x, 62.20.Dc}

\end{frontmatter}
%\tableofcontents
%\maketitle
%\newpage
%\tableofcontents
%\newpage
\section{Introduction}
\label{sec:introducion}

After the invention of the first man-made cellular material in 1930's by Munters and Tandberg~\cite{USPatentPS}, these new high-tech materials were introduced in insulation and packaging products in the daily life. Nowadays, these lightweight materials, in the form of voided-solids or foams, are widely used in variety of engineering applications~\cite{CellularSolids,SuhAM2000} such as thermal insulation, aeronautics, impact absorbing materials~\cite{Wiklo2003,Lakes2002}, electro-mechanical sensors~\cite{ReimundRev}, \etc\ 
In this paper, we present the mechanical properties (effective Young's modulus) of a layered cellular structure, in which the layers form  a truss-like micro-structure.% as shown in Fig.~\ref{fig:geometry}. 
 The macroscopic properties of the model structure are calculated using the finite element (\fe) method~\cite{Littmarck,femlab,Garboczi1995,Garboczi2002a,Garboczi2001Berryman}. The effective Young's modulus of the considered structure can be expressed by a power-law function similar to the one proposed by \citet{CellularSolids}%Eq.~(\ref{eq:powerlaw})
. Here, the expont in the power-law %$\mathfrak{n}$ in Eq.~(\ref{eq:powerlaw}) 
is expressed as a function of solid volume fraction $q$. The computed values are compared with the experimental results of thin cellular polypropylene films~\cite{WegenerAPL2003}. It is shown that there is a good agreement between numerical calculations and experimental measurements. The comparison also gives valuable information regarding micro-structure of the polypropylene films. The information could be used to optimize the materials' property for specific applications. The paper is divided into sections. In the proceeding section (\S~\ref{sec:background}), a brief review on homogenization is explained. In \S~\ref{sec:elast-prop-ideal} the geometrical and computational models are presented. In addition, the numerical results, and the influence of various geometrical parameters are presented and discussed in the same section. The numerical results are compared to those of experiments in \S~\ref{sec:comp-with-exper}. Conclusions are in \S~\ref{sec:concl-disc}.

\section{Background}\label{sec:background}
\begin{figure}[tp]
  \centering
  \includegraphics[width=5.3in]{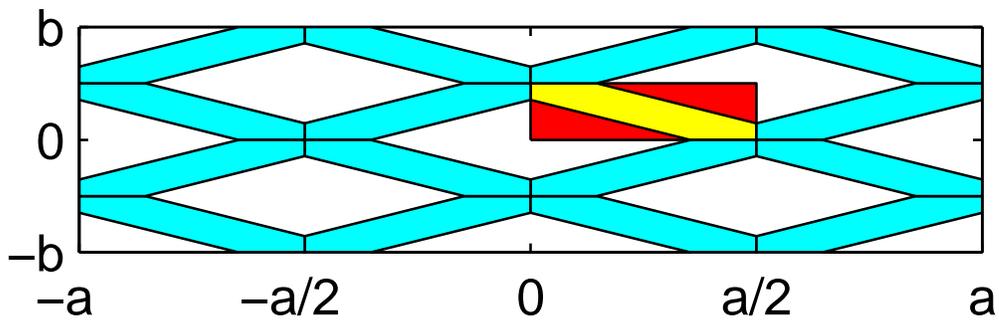}
  \caption{Considered truss-like geometry in the cellular material simulations. The shaded area represents the smallest representative unit (unit-cell).\label{fig:geometry}}
\end{figure}

The foam structure is a heterogeneous medium whose macroscopic or {\em effective} properties can be expressed as a function of its constituents.
\begin{eqnarray}
  \label{eq:effective}
  \mathfrak{P}^{\sf e}=f(q_i,\mathfrak{P}_i)
\end{eqnarray}
where $\mathfrak{P}^{\sf e}$ is the effective property sought and subscript `i' represents the properties the constituents with volume fractions $q_i$, $\sum q_i =1$ (in \citet{CellularSolids} notation $q=\rho^*/\rho$, where $\rho$ and $\rho^*$ are the density of the solid and foam, respectively). Determining $\mathfrak{P}^{\sf e}$ for  heterogeneous materials or structures or finding a representative function $f$ is an old problem in science~\cite{TorquatoBook,SahimiBookI}. The simplest forms for function $f$ are the arithmetic $f_a$ and harmonic $f_h$ averages,
\begin{eqnarray}
  f_a(q_i,\mathfrak{P}_i)&=&\sum q_i\mathfrak{P}_i \quad \text{and}   \label{eq:simplist_par}
\\
  f_h(q_i,\mathfrak{P}_i)&=&\left(\sum q_i/\mathfrak{P}_i\right)^{-1}.  \label{eq:simplist_perp}
\end{eqnarray}
These two equations are actually the bounds for mixture properties~\cite{wiener,Voigt,Reuss}. The former one, Eq.~(\ref{eq:simplist_par}),  represents a mixture with layers parallel to the applied field. The same arrangement perpendicular to the applied field direction is expressed by Eq.~(\ref{eq:simplist_perp}). For complex mixtures, \ie\ the constituents are arranged in a randomly fashion, non-trivial mixture formulas and rigorous bounds are applied~\cite{TorquatoBook,Homogenization}. Eq.~(\ref{eq:simplist_par}) is often valid for low volume fractions of constituents, however, for open-cell cellular solids, the effective Young's modulus $E^{\sf e}$ is not trivial to find due to lack of mechanical properties of the void. It was stated that $E^{\sf e}$ is related to the volume fraction of the solid material $q$ through a power-law relation~\cite{CellularSolids},
\begin{eqnarray}
  \label{eq:powerlaw}
  E^{\sf e}=\mathfrak{C}q^\mathfrak{n}\,E_s
\end{eqnarray}
where, $\mathfrak{C}$ and $\mathfrak{n}$ depend on the micro-structure, and $E_s$ is the Young's modulus of the solid. The value of $\mathfrak{n}$ lies in the range $1<\mathfrak{n}<4$, yielding a wide range of effective properties for a given solid volume fraction $q$~\cite{Garboczi2001}. \citet{CellularSolids} have presented experimental results on various cellular materials, and their observations have suggested $\mathfrak{n}=2$. Since the dependence of effective properties on micro-structure are not well understood in mixtures, the exact form of $\mathfrak{C}$ and $\mathfrak{n}$ are not known, which is an handicap to optimize and predict composite properties. However, computer simulation are prefered in such cases~\cite{TorquatoBook,CellularSolids}.

\section{Model cellular structure }
\label{sec:model-cell-struct}
\label{sec:elast-prop-ideal}

\subsection{Geometrical consideration}
\label{sec:geom-cons}

\begin{figure}[tp]
  \centering
  \includegraphics[width=5.3in]{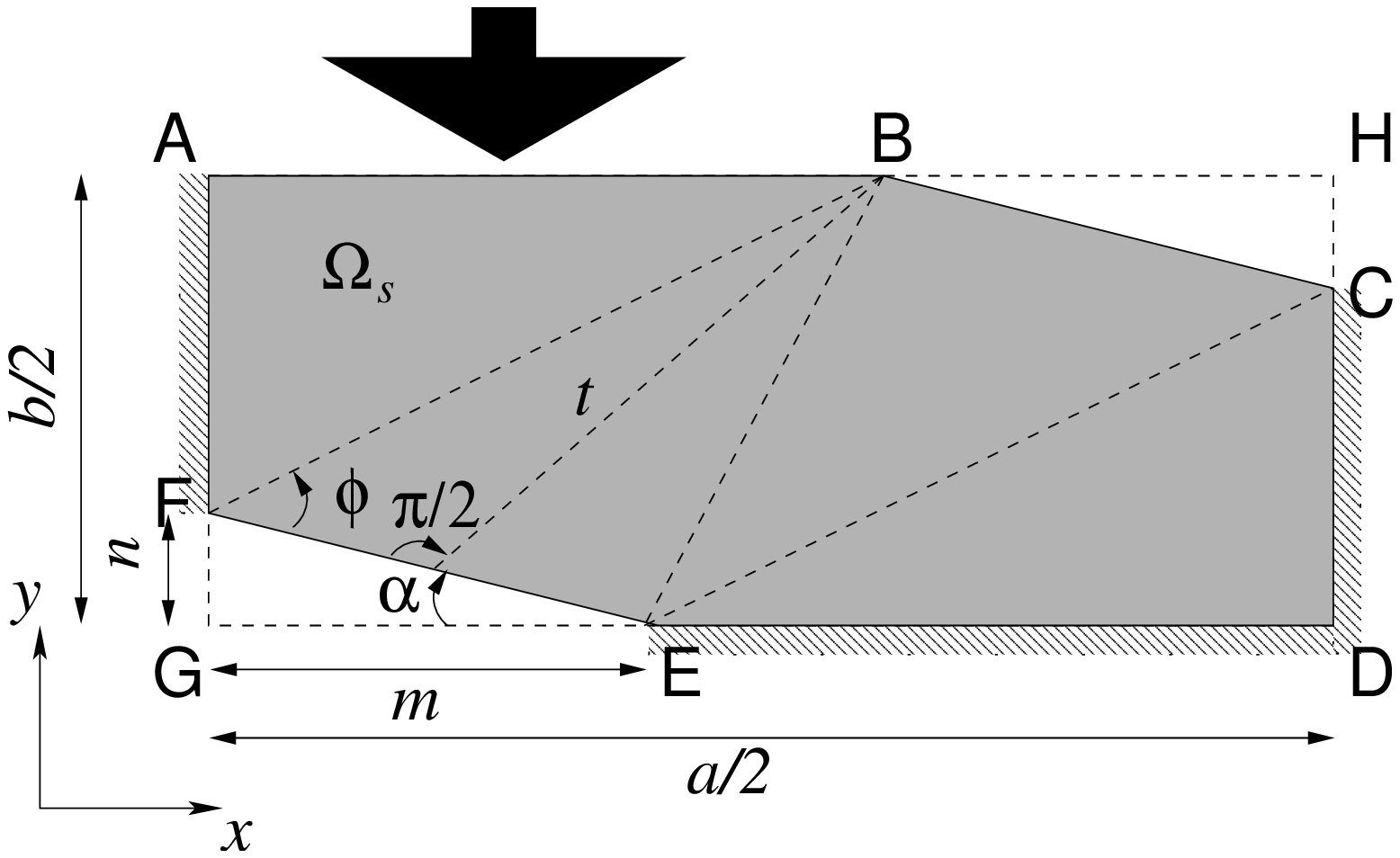}
  \caption{Unit cell (representative volume element) in the simulations. The dark-gray region (${\sf ABCDEF}$ polygon) is the solid medium, $\Omega_s$. The shaded areas represent the boundary conditions, such that the solid region is allowed to deform in the rectangular area marked by ${\sf AHDG}$. The arrow represents the force (load) applied to the edge ${\sf AB}$. The wall thickness of the solid is calculated using the angles $\widehat{\sf GEF}$ and $\widehat{\sf EFB}$, which are denoted by $\alpha$ and $\phi$, respectively. The volume fraction of the solid region $\Omega_s$, polygon ${\sf ABCDEF}$, is $ab/4-mn$.\label{fig:thicknesssketch}}
\end{figure}

 The considered two-dimensional truss-like structure is presented in Fig.~\ref{fig:geometry}. It is a symmetric geometry with solid regions and diamond-like voids centered at the triangular lattice nodes. Because of the lattice structure, the region marked between $[0,\ a/2]$ and $[0,\ b/2]$ in Fig.~\ref{fig:geometry} is used as the unit cell in the simulations. The marked region is the smallest representative part of the whole structure with $a/2$ and $b/2$ in the $x$- and $y$-directions, respectively. The geometrical model of the unit cell is illustrated in Fig.~\ref{fig:thicknesssketch}. The dark polygon between ${\sf ABCDEF}$ represents the solid medium $\Omega_s$ with wall thickness $t$. The expressions that define the geometrical parameters are as follows. 
\begin{eqnarray}
  q&=&ab/4-mn,  \label{eq:thickness} \\
  \alpha&=&\arctan (b/a),  \label{eq:alpha} \\
%  \phi&=&2\alpha\\
  m&=&\sqrt{(1-q)ab/(4\tan \alpha)}, \label{eq:m} \\
  n&=&m\tan\alpha, \label{eq:n}\\
  t(q)&=&2\sqrt{(b/2-n)^2+(a/2-m)^2}\,\sin (2\alpha) \label{eq:tofq}
%    \sqrt{(b-(1-q)b^2)^2+(a-(1-q)a^2)^2}\cos(\arctan(b/a))
\end{eqnarray}
where $q$ is the area/volume fraction of the solid medium and $\phi=2\alpha$ in Fig.~\ref{fig:thicknesssketch}. The angles $\alpha$ and $\phi$ are defined as illustrated in Fig.~\ref{fig:thicknesssketch}, and are used  to calculate the cell-wall thickness $t$. The introduced parameters $m$ and $n$ are related to the dimensions of the diamond-like void. From the above equations, one can also calculate the concentration $q$ for a given thickness $t$ and the unit-cell dimensions $a$ and $b$,
%\begin{widetext}
\begin{eqnarray}
  \label{eq:qoft}
   q(t)&=& {\frac{2\,{\left( a^2 + b^2 \right) }^{{3}/{2}}\,\tau - 
         (a\,\tau)^2 - 
         (b\,\tau)^2}
       {(a^2 + b^2)^2}},
\end{eqnarray}
%\end{widetext}
where  $\tau=t\,\csc (2\,\alpha)$. In the \fe\ analysis, %these geometrical parameters are used to calculated 
the effective Young's moduli are calculated for various $a$ and $b$ values.
 
\subsection{Numerical analysis}
\label{sec:fem}

\begin{figure}[tp]
  \centering
  \includegraphics[width=5.3in]{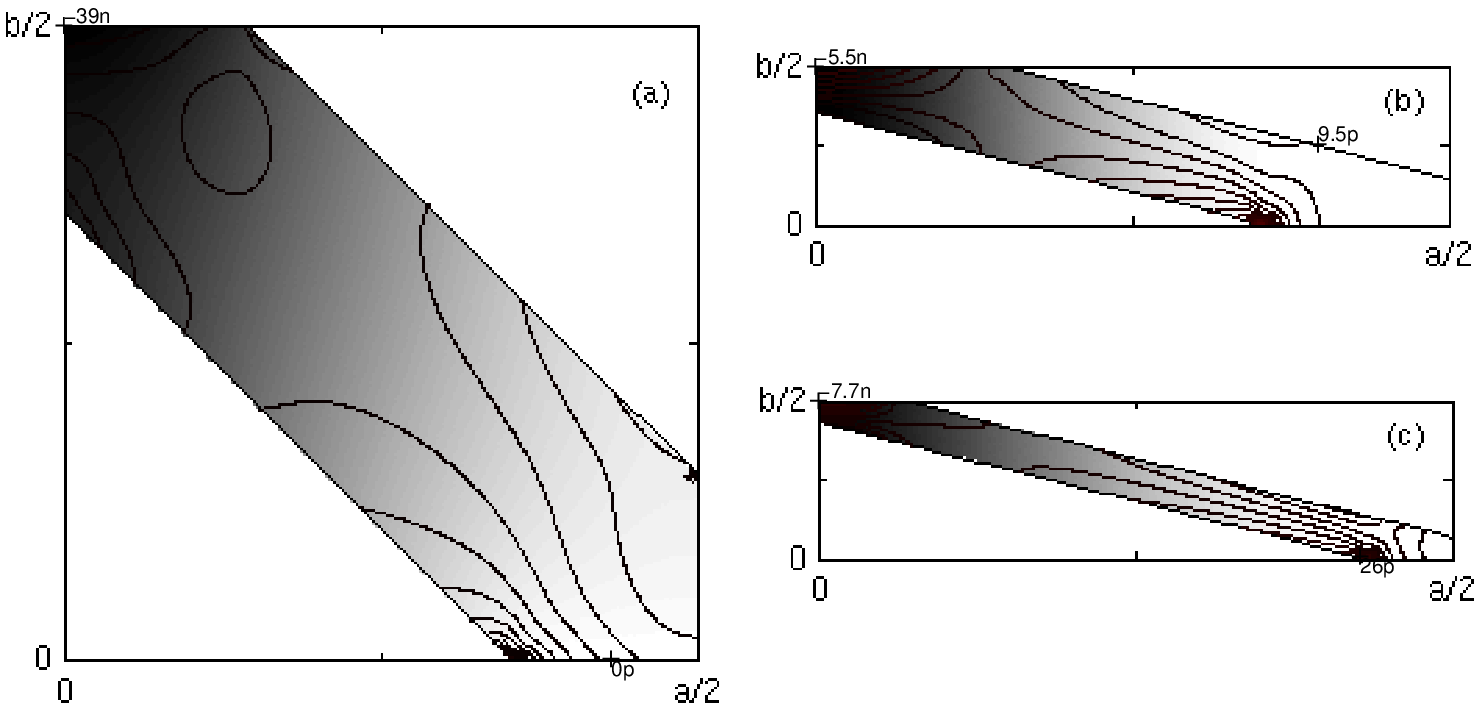}
  \caption{Global displacement (gray scale color map) and von Mises stress (contour plot) distributions for three selected structures (a) $a/b=1$ and $q=0.5$, (b) $a/b=4$ and $q=0.5$ and (c) $a/b=4$ and $q=0.25$, respectively. Darker regions show the highest deformations. The minimum and maximum deformation locations for unit applied loads are indicated in the graphs in metric units. The calculated effective Young's moduli for these structures are (a) $170\ \mega\pascal$, (b) $3.4\ \mega\pascal$ and (c) $1.0\ \mega\pascal$, respectively. \label{fig:results}}  
\end{figure}

Our \fe\ method is based on the minimization of the elastic energy in conjunction with adaptive meshing~\cite{Littmarck,FEMLAB_SMM}.  The Newton's second law in the two-dimensions, which is in reality a system of two linear equations known as Navier's equations, are solved for spatial points obtained from the \fe\ triangulation of $\Omega_s$. 
\begin{eqnarray}
  \label{eq:Navier}
  -\nabla\cdot\mathbf{T}=\mathbf{K}
\end{eqnarray}
Here $\mathbf{T}$ is the mechanical stress tensor and $\mathbf{K}$ is the load vector. The stress tensor has three components in two-dimensions, which are normal stress components $\mathbf{T}_x$ and $\mathbf{T}_y$, and the shear stress component $\mathbf{S}_{xy}$,
\begin{eqnarray}
  \label{eq:Stress}
  -\frac{\partial\mathbf{T}_x}{\partial x} 
  -\frac{\partial\mathbf{S}_{xy}}{\partial y} 
  &=&\mathbf{K_x}\\
  -\frac{\partial\mathbf{T}_y}{\partial y} 
  -\frac{\partial\mathbf{S}_{xy}}{\partial x} 
&=&\mathbf{K_y}
\end{eqnarray}
Including the Hooke's law between the stress and the elastic strain, $\mathbf{T}=c\, \nabla\mathbf{u}$, lead to a  partial differential equation expressed in the global displacements $\mathbf{u}(u,v)$. Then, Eq.~(\ref{eq:Navier}) becomes as follows
\begin{eqnarray}
  \label{eq:Navier2}
  -\nabla\cdot c \nabla\mathbf{u}=\mathbf{K},
\end{eqnarray}
where the coefficient $c$ is a function of solid materials Young's modulus and Poisson's ratio. In Fig.~\ref{fig:thicknesssketch}, the shaded areas and the arrow represent the constraints and load, respectively. The Young's modulus of the structure is  calculated from the average deformation of the edge ${\sf AB}$ in the $y$-direction $\int v\,\dd x/(1-m)$.
\begin{eqnarray}
  \label{eq:Young}
  E^{\sf e}=\frac{b\,F_0}{\int_0^{1-m} v\,\dd x}
\end{eqnarray}
\begin{table}[bp]
  \caption{Boundary conditions in the \fe. The constrained edges have $u=0$ or $v=0$ conditions for `yes' labels. All the other edges are `free' to move in corresponding directions. The applied load per unit length is presented by $F_0$.}
  \centering
  \begin{tabular*}{4.5in}{l@{\extracolsep{\fill}}rrrrrr}
%\toprule
\hline\hline
& ${\sf AB}$&${\sf BC}$&${\sf CD}$&${\sf DE}$&${\sf EF}$&${\sf FA}$\\
%\colrule
\hline
Load & $F_0$ & --- & --- & ---& --- & --- \\
x-constraint & yes & free & yes & yes & free & yes \\
y-constraint & free & free & free & yes & free & free \\
\hline\hline
%\botrule
  \end{tabular*}
  \label{tab:boundary}
\end{table}

The solid region $\Omega_s$ in Fig.~\ref{fig:thicknesssketch} is used in the simulations together with the boundary conditions presented in Table~\ref{tab:boundary}. Meshing of the computational domain is an important factor in the \fe\ analysis, we have, therefore, adapted a procedure that uses the cell-wall thickness as the meshing parameter. In the calculations, the \fe\ triangle sides are not larger than the one-tenth of the cell-wall thickness. This procedure allowed us to discretize even very thin cell walls. 

%\section{Numerical results}
%\label{sec:numer-results-disc}
\begin{figure}[bp]
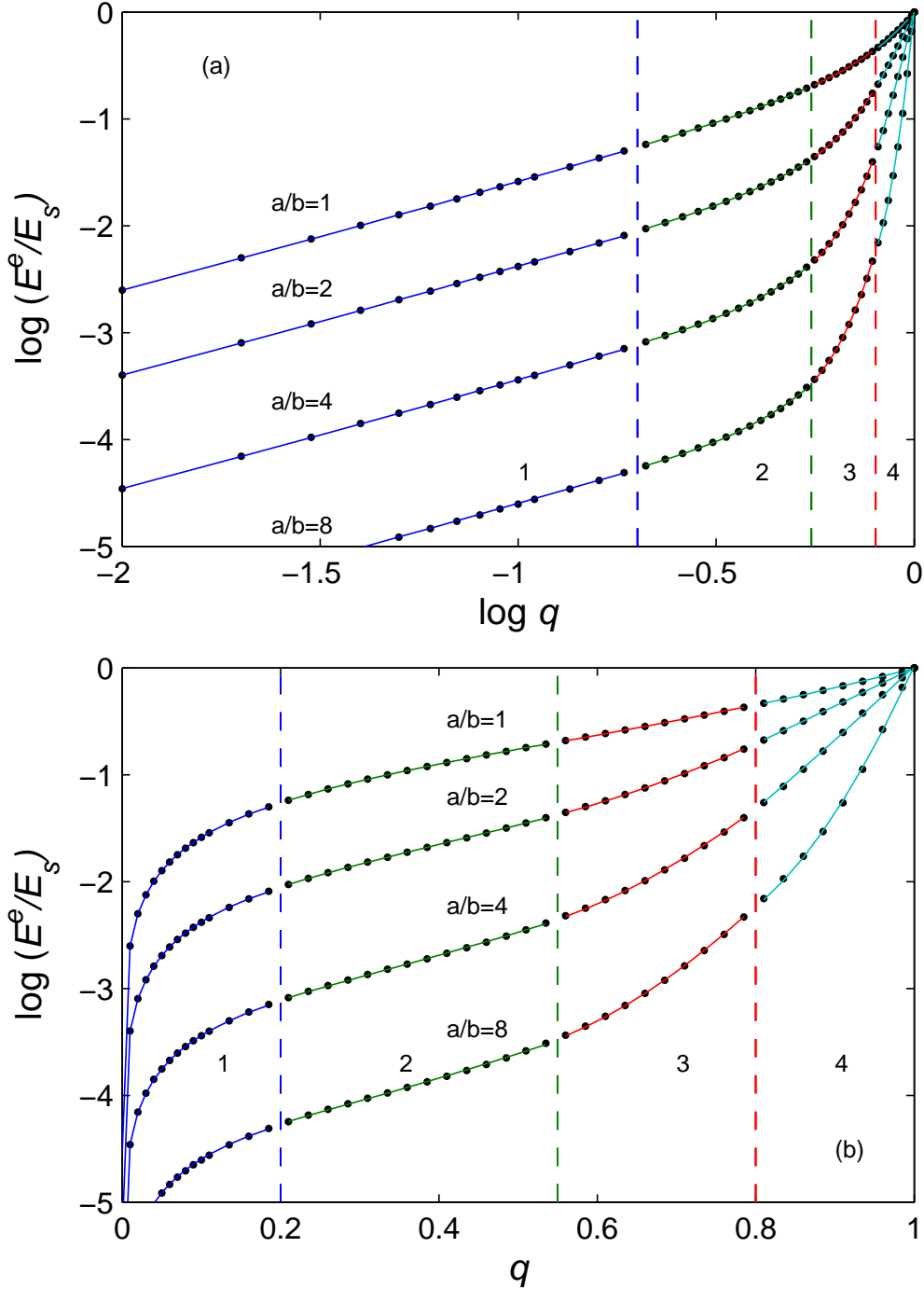

  \centering
  \includegraphics[width=5.3in]{ratio2paper_log.epsc}
  \includegraphics[width=5.3in]{ratio2paper_lin.epsc}
  \caption{Normalized effective Young's modulus $E^{\sf e}/E_s$ as a function of volume fraction of solid material $q$, (a) in log-log and (b) in semi-log plots.  Only four $a/b$ ratios are shown. The curves are divided into four sectors as separated by dashed vertical lines (\kesik) at $q=\{0.2,\ 0.55,\ 0.8\}$. The effective Young's modulus in each sectors are expressed with a power-law, Eq.~(\ref{eq:sectionfit}). The solid lines (\full) represent the power-law curves with the fitting parameters presented in Table~\ref{tab:fitting}.  \label{fig:ratiosLogLin}}
\end{figure}

%\begin{landscape}
\begin{sidewaystable}
  \caption{Fitting parameters of Eq.~(\ref{eq:sectionfit}) at different volume fraction intervals, the sectors displayed in Fig.~\ref{fig:ratiosLogLin}.\label{tab:fitting}}
  \centering
%  \begin{tabular}{5in}{rrrrrrrrrrrrr}%{r@{\extracolsep{\fill}}rrrrrrrrrrrr} 
  \begin{tabular*}{6.8in}{@{\extracolsep{\fill}}rcccccccccccc} 
    \hline  \hline
    & \multicolumn{3}{c}{$q\le0.20$} & \multicolumn{3}{c}{$0.20\le q\le0.55$} &\multicolumn{3}{c}{$0.55\le q\le0.80$} &\multicolumn{3}{c}{$q\ge0.80$} \\
%\cline{2-4}
%    & \crule{3} &\crule{3} &\crule{3} &\crule{3} \\
$a/b$ & $\log \mathfrak{a}_1$ &$ \mathfrak{b}_1$ &$\mathfrak{c}_1$ &$\log \mathfrak{a}_2$ &$\mathfrak{b}_2$ &$\mathfrak{c}_2$ &$ \mathfrak{a}_3$ &$\mathfrak{b}_3$ &$\mathfrak{c}_3$ &$ \mathfrak{a}_4$ &$\mathfrak{b}_4$ &$\mathfrak{c}_4$ \\
\hline
1 & $-0.69$ & $-0.67$ & 0.96 & $-0.14$ & 1.53 & 1.30 &0.93 & ~2.76  & ~~$ 1.04$ &0.99 & ~4.61  &~$-0.14$ \\ 
2 & $-1.50$ & $-0.77$ & 0.96 & $-0.46$ & 3.59 & 1.55 &1.08 & ~8.96  & ~~$ 0.49$ &1.00 & ~4.66  &~~~$ 3.59$ \\ 
3 & $-2.10$ & $-0.77$ & 0.96 & $-0.82$ & 4.70 & 1.68 &1.13 & 14.21 & $-0.21$ &1.01 & ~5.81  &~~~$ 6.02$ \\ 
4 & $-2.56$ & $-0.79$ & 0.96 & $-1.15$ & 5.32 & 1.74 &0.95 & 17.62 & $-0.71$ &1.03 & 12.91 &~~~$ 3.45$ \\ 
5 & $-2.93$ & $-0.77$ & 0.96 & $-1.45$ & 5.63 & 1.77 &0.70 & 19.63 & $-0.97$ &1.05 & 24.65 &~$-3.12$ \\ 
6 & $-3.24$ & $-0.80$ & 0.96 & $-1.71$ & 5.87 & 1.79 &0.49 & 20.85 & $-1.11$ &1.06 & 38.87 &$-11.99$ \\ 
7 & $-3.49$ & $-0.77$ & 0.96 & $-1.94$ & 6.01 & 1.81 &0.34 & 21.64 & $-1.19$ &1.06 & 53.98 &$-21.87$ \\ 
8 & $-3.72$ & $-0.79$ & 0.96 & $-2.14$ & 6.12 & 1.82 &0.23 & 22.13 & $-1.22$ &1.05 & 68.86 &$-31.84$ \\ 
9 & $-3.93$ & $-0.84$ & 0.95 & $-2.32$ & 6.22 & 1.82 &0.17 & 22.51 & $-1.24$ &1.02 & 83.45 &$-41.81$ \\ 
10& $-4.11$ & $-0.84$ & 0.95 & $-2.50$ & 6.20 & 1.83 &0.12 & 22.70 & $-1.21$ &0.99 & 97.34 &$-51.42$ \\
\hline\hline
\end{tabular*}
\end{sidewaystable}
%\end{landscape}

In the calculations the Young's modulus of the solid is $E_s=1\ \giga\pascal$~\cite{matweb}, and its Poisson's ratio is $\nu_s\approx0$. The effective Young's moduli of the structures are practically independent of the $\nu_s$, which has also been observed previously~\cite{Garboczi1992,Garboczi2002a}. 

\subsection{Illustrative examples}

Three simulations are illustrated in Fig.~\ref{fig:results}. The gray scale map and contours represent the global deformation and von Mises stress distributions $\mathbf{T}_{\text{vM}}$,
\begin{eqnarray}
  \label{eq:vonMises}
  \mathbf{T}_{\text{vM}}=\sqrt{\mathbf{T}_x^2+\mathbf{T}_y^2-\mathbf{T}_x\,\mathbf{T}_y+3\,\mathbf{S}_{xy}}
\end{eqnarray}
In the figures, the dimensions of the representative units are varied in order to display the change in the stress and displacement distributions for selected $a/b$ and $q$ values. For given structural parameters $a/b$ and the volume fraction of the solid $q$ yield distint global displacements and von Mises stress. The effective Young's moduli for these structures are $170\ \mega\pascal$, $3.4\ \mega\pascal$ and $1.0\ \mega\pascal$ for Figs.~\ref{fig:results}a, \ref{fig:results}b and \ref{fig:results}c, respectively. The positions where the $\max$ and the $\min$ deformations are observed are also indicated in the figures. Depending on $a/b$ and $q$, the bending of the beam-like cell-wall structure is peculiar, and the cell-wall thickness is an important factor for this behavior. %To sum up, the unit-cell dimensions $a/b$ and volume fraction of solid $q$ yield distinct effective Young's moduli $E^{\sf e}$.

\subsection{Effective properties of the cellular structure}
\label{sec:effect-prop-cell}

The effective Young's moduli, calculated from the deformations, for various $a/b$ ratios are shown  as log-log and semi-log plots in Fig.~\ref{fig:ratiosLogLin}. In the figure, four of the simulation results are illustrated, $a/b=\{1,\ 2,\ 4,\ 8\}$. At low concentrations of the solid ($q\,\le\,0.20$), the effective Young's modulus $E^{\sf e}$ is a linear function of the volume fraction in the log-log representation [as in Eq.~(\ref{eq:powerlaw})]. However, the constant $\mathfrak{C}$ is related to the unit-cell dimensions $a$ and $b$. When the volume fraction of the solid is increased there is no simple relation between $E^{\sf e}$ and $q$, which is illustrated in Figs.~\ref{fig:ratiosLogLin}a and \ref{fig:ratiosLogLin}b. In the intermediate low concentrations $0.20\,\le\,0.55$, the dependence of $E^{\sf e}$ is slightly semi-logarithmic with similar slopes but the shifts are different related to the unit-cell dimensions (Fig.~\ref{fig:ratiosLogLin}b). At higher concentration of solid, the behavior of $E^{\sf e}$ is still a power-law dependence on $q$. It is non-trivial to find an homogenization formula that can express the effective Young's modulus $E^{\sf e}$ as a function of whole range of solid volume fraction ($0<q\le1$) and the unit-cell dimensions for the truss-like structure. However, by dividing the $q$-axis into four separate sectors. 
%the Young's modulus of the structures can be described by a power-law similar to Eq.~(\ref{eq:powerlaw}). 
The facing sectors are separated at $q=\{0.20,\,0.55,\,0.80\}$, and the dashed vertical lines (\kesik)  in Fig.~\ref{fig:ratiosLogLin} are drawn to represent the borders. We are able to represent the effective Young's modulus of the truss-like structure with a power-law similar to Eq.~(\ref{eq:powerlaw}) in each sector when we modify the exponent $\mathfrak{n}(q)$, $\mathfrak{n}=\mathfrak{a}\,q+\mathfrak{b}$.   The proposed homogenization equation is then expressed as  
\begin{eqnarray}
  \label{eq:sectionfit}
    E^{{\sf e}}_{i}(q)=[\mathfrak{a}_iq^{(\mathfrak{b}_i q+\mathfrak{c}_i)}]E_{s} \quad \text{with} \quad i={1,2,3,4}
\end{eqnarray}
where $\mathfrak{a}_i$, $\mathfrak{b}_i$ and $\mathfrak{c}_i$ are the fitting parameters, and $i$ is the sector number. The values of these parameters obtained from a curve-fitting procedure for each sector are presented in Table~\ref{tab:fitting}. The resulting curves are shown in Fig.~\ref{fig:ratiosLogLin} with solid lines (\full). It is remarkable that only $\mathfrak{a}_4$, $\mathfrak{b}_1$ and $\mathfrak{c}_1$ show very slight dependence--they are nearly constant--on the dimensions of the representative unit $a$ and $b$, and $\mathfrak{b}_1\,q + \mathfrak{c}_1\approx1$. Last but not least, the curves are shifted down with increasing $a/b$, because of the increasing anisotropy in the structure. %Since the number of data point are limited as $q\rightarrow1$, Eq.~(\ref{eq:sectionfit}) with $\mathfrak{a}_4$, $\mathfrak{b}_4$ and $\mathfrak{c}_4$

The behavior of effective properties are of importance at low concentrations of solid material for technological applications, since the material would be light (the density of solid would be low)~\cite{SuhAM2000,CellularSolids}. For this reason, we first pay our attention to sector 1. In this sector, the exponent term in Eq~(\ref{eq:sectionfit}) is lower than 2 and it is approximately 1, which coincides with the arithmetic averages, Eq.~(\ref{eq:simplist_par}). Moreover, when $\mathfrak{n}\approx 1$, it is an indication that the cell-wall stretching is a dominant behavior~\cite{CellularSolids,Green1985}. Here, it is shown that for two dimensions this exponent is also 1, which clearly illustrates that we simulate the cell-strecthing behavior for $q<0.2$. Although, $E^{\sf e}/E_s\propto q$, one should adopt an {\em effective volume fraction}, $\mathfrak{a}_1$ in our case, as a function of structure dimensions $a$ and $b$. A sum of an exponential and a power-law was sufficient to describe the behavior of $\mathfrak{a}_1$,
\begin{eqnarray}
  \label{eq:a_1}
  \mathfrak{a}_1(a,b)&=&A\,\exp[B\,(a/b)]+C\,(a/b)^B
\end{eqnarray}
Here the constants $A=-16.84$ $B=-3.845$ and $C=0.565$, respectively. As the ratio of the cell dimension increases $a/b\,\rightarrow\,\infty$, $\mathfrak{a}_1(a/b)\,\rightarrow\,0$. As a result, for low concentrations of solid, the effective Young's modulus of the truss-like structure in two-dimensions can be expressed as
\begin{eqnarray}
  \label{eq:lowconcentration}
  E^{{\sf e}}_{i}(q\le0.2) \approx  \{A\,\exp[B\,(a/b)]+C\,(a/b)^B\}\,q
\end{eqnarray}

In sectors 2, 3 and 4, the relations between fitting parameters and $a/b$ are non-trivial. Although the fitting parameters in Table~\ref{tab:fitting} indicate clear dependence on the unit-cell dimensions $a/b$, expressions in the form of Eq.~(\ref{eq:a_1}) are not sought. %In \S~\ref{sec:effect-prop-cell}, the effective Young's modulus should approach to $E_s$ linearly as $q\,\rightarrow\,1$ similar to the very low concentrations $q\,\lll\,1$.

%It is striking that the exponent stated by \citet{CellularSolids} for open-cell cellular structures in three-dimensions is $\mathfrak{n}=2$. 

As a note, if we consider the similarity to the Bruggeman symmetrical~\cite{Bruggeman1935,Landauer1978} formula or general Wiener formula~\cite{Tuncer2001a,greffePier6,Pier6}, which use a dimensionality parameter $d$ in the effective permittivity expressions, the pre-exponent $\mathfrak{a}_1$ here is related to the internal structure of the material at low concentrations of solid. \citet{Tuncer2001a} have stated that the shape parameter in dielectric mixtures could in principle be used as the dimensionality. In the present problem, the shape parameter is hidden in the $a/b$, such that when $a/b\rightarrow \infty$ the structure is effectively a lateral layered structure, and as stated in \S~\ref{sec:introducion}, the void does not have a mechanical property ($E_v=0$), therefore $E^{\sf e}\,\rightarrow\,0$ even for high volume fractions of solid medium. At high $a/b$ values (for anisotropic structures), the effective load carrying volume of the solid is small. This is also observed in Fig.~\ref{fig:ratiosLogLin} in sectors 3 and 4 on which $E^{\sf e}/E_s$ drops fast with increasing $a/b$. And when $a/b\rightarrow1$, the change in $E^{\sf e}$ is not as drastic as $a/b>1$ cases as $q$ decreases from $1$ to $0.6$, due to the effective load carrying sections of the structure are larger in volume fraction.

\subsection{Influence of the cell-wall thickness}
\label{sec:cell-wall-thickness}
\begin{figure}[tp]
  \centering
  \includegraphics[width=5.3in]{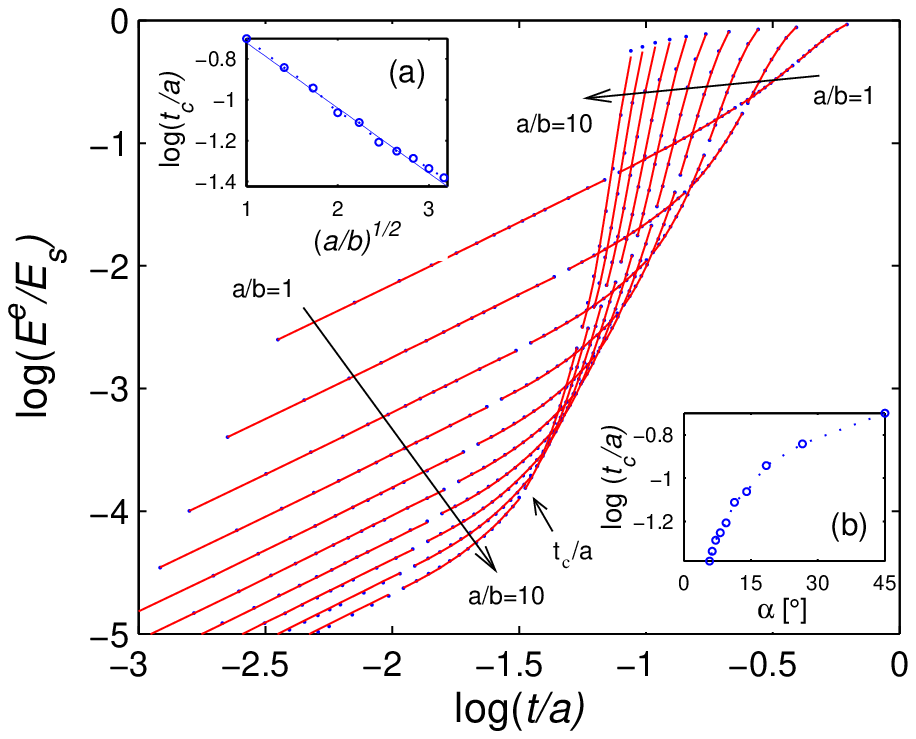}
  \caption{Normalized effective Young's modulus $E^{\sf e}/E_s$ as a function of normalized cell wall thickness $t/a$. The critical thickness for $a/b=10$ is marked with $t_c$, and the inset (a) displays the linear dependence of $\log\,(t_c/a)$ on square root of dimensions ratio $(a/b)^{1/2}$, the solid line (\full) is $\log(t_c/a)=0.32\,(a/b)^{1/2}-0.4$. The inset (b) is the same dependence as a function of the angle $\alpha$ defined in Eq.~(\ref{eq:alpha})\label{fig:tvsE}}
\end{figure}
\citet{CellularSolids} has shown that the mechanical properties of foams could be scaled when the cell-wall thickness was much smaller than the relative cell size ($q\ll1$), 
\begin{eqnarray}
  \label{eq:thicknessscale}
  E^{\sf e}=\mathfrak{B}\,(t/b)^\mathfrak{m}\,E_s,
\end{eqnarray}
where $\mathfrak{B}$ and $\mathfrak{m}$ are scaling parameters, and $t$ and $b$ are the cell-wall thickness and unit-cell height for $a/b=1$. They have stated that the value of $\mathfrak{m}$ is 3 and 4 for open- and closed cell foams, respectively.

 The effective Young's modulus as a function of the cell-wall thickness is illustrated in Fig.~\ref{fig:tvsE} for 10 different unit-cell dimensions. The behavior is obtained by converting the relation in Eq.~(\ref{eq:sectionfit}) as a function of cell-wall thickness by substituting Eq.~(\ref{eq:qoft}) in Eq.~(\ref{eq:sectionfit}). It is clearly indicated that the thicker walls yield greater mechanical stiffness because of the higher volume fraction of the solid $q$~\cite{Barrett1994}. This result is also discussed in \S~\ref{sec:effect-prop-cell}. Structures with thin walls and low $q$ solid yield very low values of elastic moduli, and $E^{\sf e}/E_s\propto (t/a)$.

There is a critical knee-point $t_c$ at which the power-law behavior of $E^{\sf e}$ changes its exponent from a lower value to a higher one. As we approach the critical thickness from high thickness values the physical nature of the truss-like structure changes the stress field distribution, and the deformation is altered. The position of the critical thickness is calculated from a similar expression as in Eq.~(\ref{eq:sectionfit}), however, only two regions are considered, $t/a<t_c/a$ and $t/a>t_c/a$--the regions are separated by $t_c/a$. The logarithm of the critical thickness against $(a/b)^{1/2}$ indicate linear dependence on the unit cell dimensions, which is shown in the inset (a) of Fig.~\ref{fig:tvsE} and $E^{\sf e}/E_s\propto (t/a)^{4.8\sqrt(a/b)-3.2}$. If a similar relation as Eq.~(\ref{eq:thicknessscale}) is  considered for $t/a>t_c/a$, the exponent in Eq.~(\ref{eq:thicknessscale}) has an $a/b$ dependence, $\mathfrak{m}=\mathfrak(a/b)$. Finally, since $a/b$ is related to the angle between the truss-like surfaces, for a given volume fraction $q$, the soft and stiff structures can be tailored by varying the unit-cell dimensions. This is illustrated in the inset (b) in Fig.~\ref{fig:tvsE}. At large angle values $\alpha\gtrsim20\degree$, the critical thickness changes slowly compared to the values at low angles $\alpha\lesssim15\degree$. %If $q$ is sought for the critical thickness values, it is observed that as $\a/b \rightarrow10$ $q\rightarrow \sim1/3$ as presented in Table~\ref{tab:criticalthickness}. For structures with $a/b>5$, it is  

%\begin{table*}[tp]
%  \caption{Critical cell-wall thicknesses and resulting volume fraction of solid medium for various unit-cell dimensions.}
%  \centering
%  \begin{tabular*}{6in}{c@{\extracolsep{\fill}}cccccccccc}
%\hline\hline
%$a/b$ & $1$&$2$&$3$&$4$&$5$&
%          $6$&$7$&$8$&$9$&$10$\\
%\colrule
%$t_c/a$ & 0.200 & 0.144 & 0.114 & 0.087 & 0.077 & 0.062 & 0.056 & 0.052 & 0.046 & 0.041  \\ 
%$q$ & 0.485 & 0.540 & 0.592 & 0.586 & 0.633 & 0.612 & 0.635 & 0.660 & 0.659 & 0.660 \\
%\botrule
%  \end{tabular*}
%  \label{tab:criticalthickness}
%\end{table*}

\section{Comparison of numerical results with experiments}
\label{sec:comp-with-exper}

%\subsection{Cellular polypropylene}
\label{sec:cell-polypr}
\begin{figure}[tp]
  \centering  \includegraphics[width=5.3in]{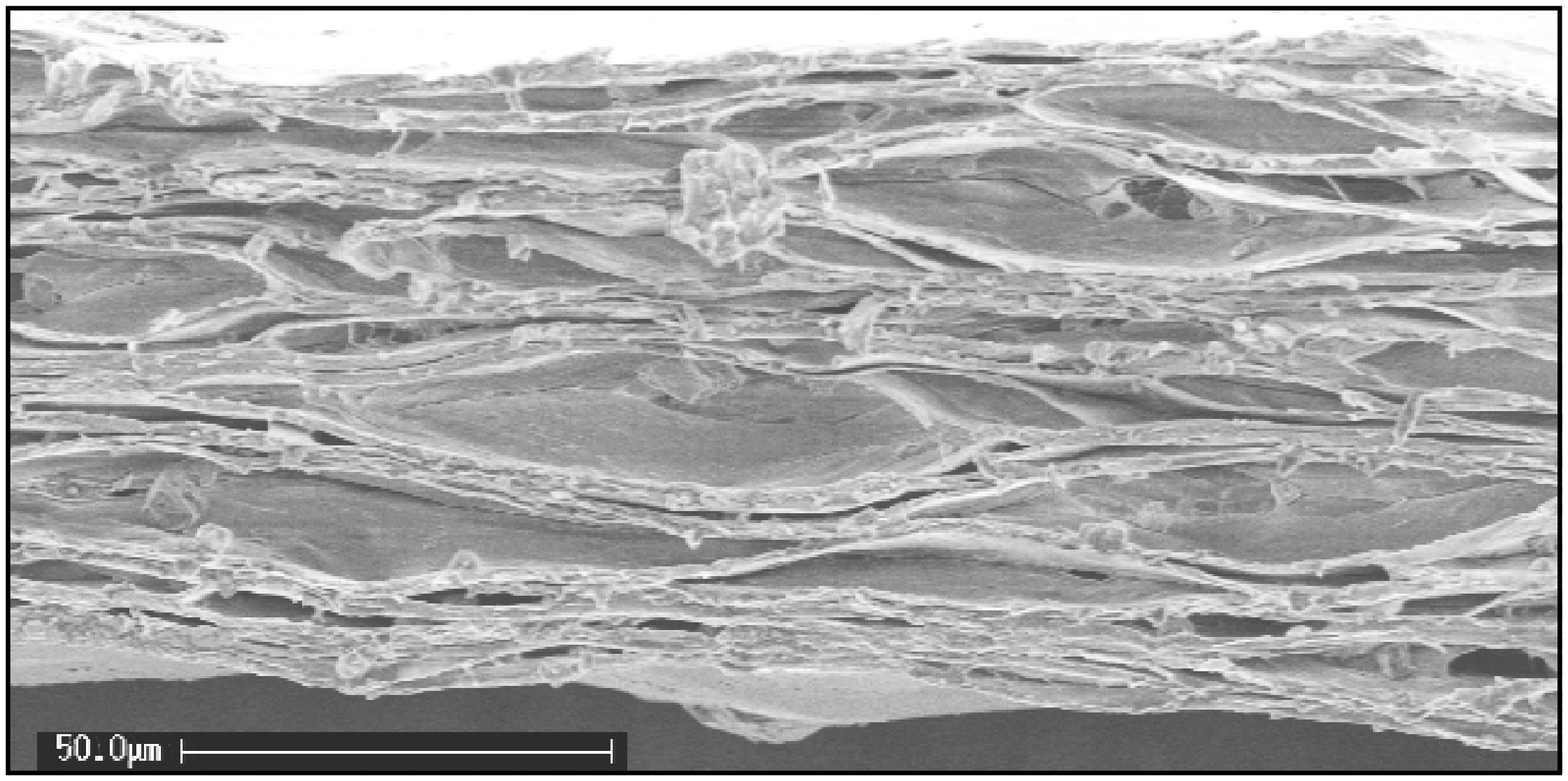}
  \caption{\label{PPreal} Cross-sectional scanning electron microscope image of a cellular polypropylene film. The dark regions are air-voids and bright spots are polymer. The film was inflated with $25\,\bbar$ at $90\,\celsius$ from a $0.73\,\gram\per\centi\cubic\meter$ and $37\,\micro\meter$ thick sample. The density of the sample is $0.44\,\gram\per\centi\cubic\meter$. }
\end{figure}

 In Fig.~\ref{PPreal}, the scanning electron microscope cross-sectional view of a cellular \pp\ film is shown. The dark areas indicate the void regions, and the bright regions are the solid polypropylene, $\Omega_s$. The original cellular \pp\ films were supplied from VTT Processes, Finland. They were $37\ \micro\meter$ thick and stretched~\citep[the detailed description of the material can be found in Ref.][]{Lekkala2000}. The films were inflated by means of a pressure treatment in order to vary the dimensions of the cells~\cite{PaajanenISE2002,WegenerAPL2003,Wegener-CEIDP03,Hillenbrand-CEIDP03}. %\citep[the procedure is briefly described in Ref.][]{WegenerAPL2003}. 
The treatment not only influences the shape and the size of the cells, but also the relative density (the solid volume fraction $q$).   Several samples were prepared and inflated for each individual measurement. The effective Young's modulus of the samples has been estimated from dielectric measurements~\cite{AxelReview}. % but not from the resonance itself in each data.%. This method is widely used in piezoelectricity measurements

The micro-structure of the polypropylene films indicate a cell structure similar to the regular geometry used in the \fe\ analysis in \S~\ref{sec:geom-cons}, but the cell-walls do not show a sharp connection points (corners) as in Fig.~\ref{fig:geometry} (or the nodes {\sf E} and {\sf F}, and {\sf B} and {\sf C} in Fig.~\ref{fig:thicknesssketch}). The structure show high anisotropy, and the cell-dimensions are not equal, $a/b\ne1$. Moreover, the size distribution of voids in the polypropylene are not constant but distributed--the structure is irregular. Yet, comparison with the simple truss-like model would yield the most probable or in other words the average void size and the cell wall thickness for material design purposes.

%\subsection{Elastic properties of cellular polypropylene}
%\label{sec:elast-cell-polypr}
\begin{figure}[tp]
  \centering
  \includegraphics[width=5.3in]{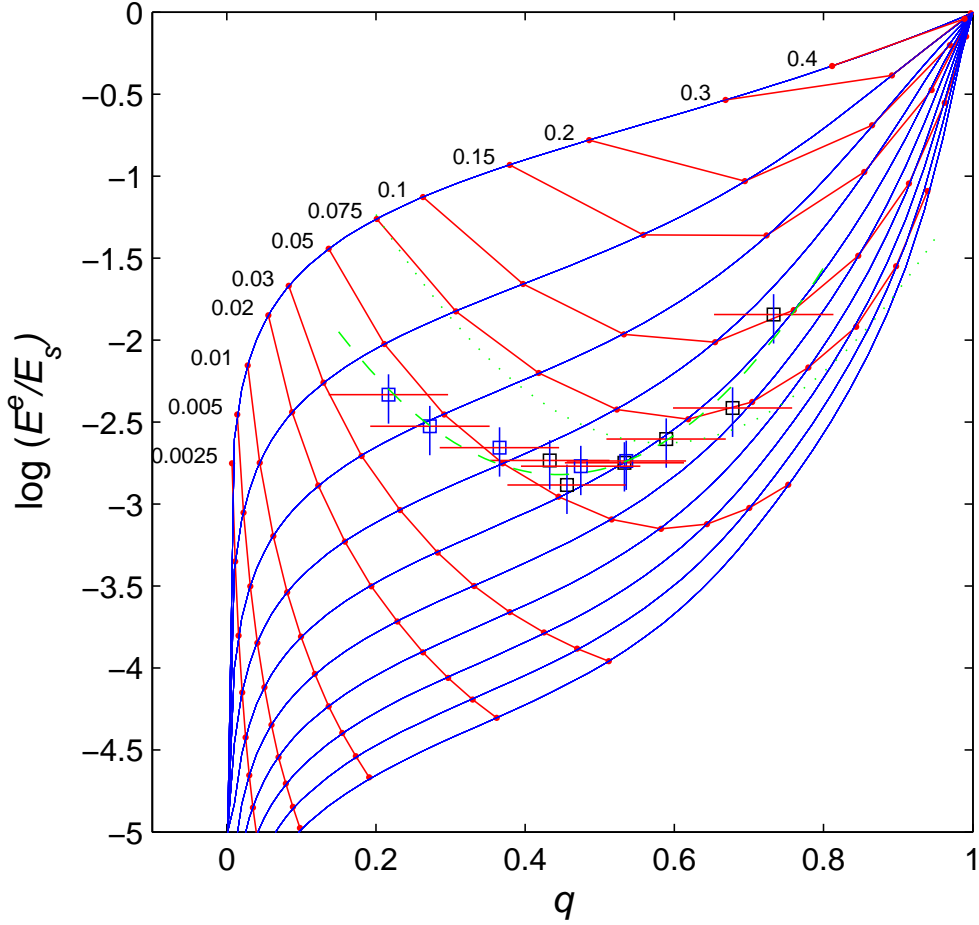}
  \caption{Normalized effective Young's modulus $E^{\sf e}/E_s$ as a function of solid volume fraction $q$. The experimental data are illusrated with symbols ($\Box$). Iso-thickness lines are also shown with thick lines, and corresponding normalized thickness values are indicated for each line. The dashed (\kesik) and dotted (\dotted) lines represent quadratic relations for the experimental ($\log E^{\sf e}/E_s = 10\,x^2-8.9\,x-0.84$) and the iso-thickness line $t/a=0.075$ ($\log E^{\sf e}/E_s = 9.4\,x^2-11\,x+0.59$), respectively.\label{fig:qtvsE}}
\end{figure}

To display the utility of the \fe\ modeling, the simulations are compared with experimental data of the cellular \pp\ films. The numerical data  and the experimental results are ploted against the volume fraction of the solid (polypropylene) $q$ in Fig.~\ref{fig:qtvsE}. Iso-thickness lines from the simulations are also displayed in the figure. The experimental data are shown with symbols ($\Box$) and error bars (the samples were thin ($\lesssim100\ \micro\meter$) and soft, therefore, the uncertainty in the thickness determination leads the error bars).   The Young's moduli of polypropylene samples were normalized by the value for the extrusion grade polypropylene~\cite{matweb}, $E_s\approx1\,\giga\pascal$. The measured values were between $1\,\mega\pascal$ and $10\,\mega\pascal$ depending on the density of the samples. The logarithm of the data indicate a quadratic dependence on the volume fraction 
\begin{eqnarray}
  \label{eq:quadratic}
  \log E^{\sf e}/E_s = 10\,x^2-8.9\,x-0.84.   
\end{eqnarray}
The dashed line (\kesik) corresponds this expression in the figure. In order to show the agreement between the numerical and experimental behavior, we apply the same  curve fitting procedure to the iso-thickness line at $t/a=0.075$. Similar to the experimental data, the considered iso-thickness line can be expressed by a quadratic expression, shown as the dotted line (\dotted) in the figure. This illustrates that the change in the behavior of stiffness might be due to the alteration of the average cell dimensions.

However, a detailed analysis of the experimental data and the numerical simulations indicates that by inflating the samples first the cell-wall thickness  is altered--cell-wall stretching~\cite{CellularSolids,Green1985}. The experimental data on $q>0.4$ can be expressed by the truss-like structure with cell dimensions $a/b\approx5$. When $q<0.4$, the data bend, and the sign of $\dd E^{\sf e}/\dd q$ becomes negative. The Young's moduli of the samples slightly follow the iso-thickness line $t/a=0.05$ for $q<0.4$. This shows that the cell-wall stretching is decreased. 

The critical thickness for polypropylene is around $0.05$. The samples become stiffer at low solid concentrations. The extrapolation of the data using Eq.~(\ref{eq:quadratic}) yields the thinnest cell-wall thickness $t/a\approx0.03$ for the polypropylene films.

In overall, the experimental data lie inside the region marked by $2< a/b \lesssim 6$ and normalized thickness $0.03\le t/a \le 0.12$.  Since, monitoring and measuring the Youngs'moduli of the samples in-situ are not presently possible during inflation, we have not been able to indicate the experimental traces of the iso-thicknesses.  However, the results indicate that there is a good aggrement between the behavior of the inflated samples and the model structure used in the \fe\ analysis.

\section{Conclusions}
\label{sec:concl-disc}
In this paper, we present the numerical calculation of effective Young's modulus for a two-dimensional truss-like structure. A new homogenezation formula is proposed, which expresses the effective properties of the system in a broder concentration range. In this procedure the concentration axis is divided into regions. The expression yields a log-log dependence of effective Young's modulus on solid concentration $q$ when $q<0.2$. Similar kind of behavior has also been observed  for foams previously~\cite{CellularSolids,Green1985}.  The numerical methods were compared with experimental data of thin cellular polypropylene film. The comparison shows that once the density of the inflated samples are known, we are able to estimate the average cell dimension and cell-wall thickness. The samples indicated that cell-wall stretching is important at solid concentrations over 0.4, which yield cell dimensions $a/b\approx5$. At high solid concentration, most of the solid material are mechanically passive yielding low effective Young's modulus. As the samples were inflated even further ({\em i}) the cell-dimensions are altered, ({\em ii}) stiffer structures with $a/b\approx2$ are obtained, and ({\em iii}) the cell-stretching behavior is decreased. As a final remark, %when the cell-wall thickness is taken into consideration, 
the manufacturing process of the cellular polypropylene films picks out particular cell-wall thicknesses. %Last but not least, if the density of the original sample is known, one can estimate the mechanical properties after the inflation process.

%It is necessary to improve the present approach to more complex two- and even to three-dimensional models as in Refs.~\citealp{CellularSolids}, \citealp{Garboczi2001} and \citealp{Garboczi2002}.   
%\section*{References}
%\newpage
\subsection*{Acknowledgement}
We thank  Mr. W. Wirges for his technical assistance in the experiments.
\bibliography{newref}
\bibliographystyle{unsrtnat}
\end{document}